# Systematic Mapping Study On Security Threats in Cloud Computing

Carlo Marcelo Revoredo da Silva, José Lutiano Costa da Silva, Ricardo Batista Rodrigues, Leandro Marques do Nascimento, Vinicius Cardoso Garcia
Informatics Center, CIn
Federal University of Pernambuco, UFPE
Recife, Brazil
{cmrs, jlcs2, rbr, lmn2, vcg}@cin.ufpe.br

*Abstract*— *Today, Cloud Computing is rising strongly, presenting itself to the market by its main service models, known as IaaS, PaaS and SaaS, that offer advantages in operational investments by means of on-demand costs, where consumers pay by resources used. In face of this growth, security threats also rise, compromising the Confidentiality, Integrity and Availability of the services provided. Our work is a Systematic Mapping where we hope to present metrics about publications available in literature that deal with some of the seven security threats in Cloud Computing, based in the guide entitled "Top Threats to Cloud Computing" from the Cloud Security Alliance (CSA). In our research we identified the more explored threats, distributed the results between fifteen Security Domains and identified the types of solutions proposed for the threats. In face of those results, we highlight the publications that are concerned to fulfill some standard of compliance.*

## I. INTRODUCTION

Cloud computing (CC), is a new concept that has the goal to make computational resources available as services on demand, in a short period of time and usage based cost. Cloud Computing is presented in three strategic business models: Infrastructure-as-a-Service (IaaS), Platform-as-a-Service (PaaS), and Software-as-a-Service (SaaS). The aim of cloud computing models (CCM) is to cut operational costs and, more important, to allow IT departments to focus on strategic projects instead of being concerned only in keeping their datacenters working [Velte et al, "Cloud Computing, A Practical Approach", McGraw-Hill Osborne Media, 1st edition, 2009]. With such benefits, CC has become a world trend and an area of strong investments. According to Gartner [2], the financial investment on CC in 2016 will have a Global Compounded Annual Growth Rate (CAGR) of: IaaS: 41%, PaaS: 26.6% and SaaS: 17.4% in 2016 [2]. In this scenario, there is growing concern in relation to the security of services provided. In the same Gartner survey, the category Management and Security will have a CAGR of 27.2%. The security policies are present in the Quality of Service term (QoS), specified in the Service Level Agreement (SLA).

In fact, many solutions are being proposed in literature. However the resulting problems from Security Threats to Cloud Computing Models (STCCM) are even newer. Those threats compromise the CIA of the resources provided. Currently we may consider seven different threats: #1 Abuse and Nefarious Use of Cloud Computing, #2 Insecure Interfaces and APIs, #3 Malicious Insiders, #4 Shared Technology Issues, #5 Data Loss or Leakage, #6 Account or Service Hijacking and #7 Unknown Risk Profile [3]. One of the reasons why those threats are so challenging is because in cloud computing the computational resources are the result of homogeneous data centers. This characteristic means that there is not an individual and proper management for each data center, making harder the adoption of an efficient security model that fulfills the specifications of the security policies [4].

Currently there are several organizations motivated research in order to minimize STCCM, for example, the Cloud Security Alliance (CSA), an organization that arose in face of those concerns. But we may also mention other organizations, such as the National Institute of Standards and Technology (NIST), the European Network and Information Security Agency (ENISA), the OWASP [51] Foundation with the project OWASP-Cloud, and the Computer Emergency Response Team (CERT). One of CSA's lines of research is precisely the compilation of a guide defining STCCM.

An approach technique to detect deficiencies in a given theme is to present a Systematic Mapping (SM) of literature. A Systematic Mapping is a revision with a given degree of amplitude of primary studies, with the goal to identify evidences and lacunae that remain in the current literature, providing a systematic focus for future revisions [5]. The result is a general overview of the researched area, where is possible to evidence the results acquired over time, therefore, identifying trends [6] [7].

The aim of this study is to benefit from SM techniques and analyze works available in literature that deal with threats and elaborate metrics with the goal to identify which threats are being more considered in literature and what kinds of solutions are being proposed. In consequence, we pretend to observe which ones of those works care to comply with some compliance standard, that from our point of view we consider able to reduce the problem related to the transparency between the security of the offered service and the client using it. Our work is structured as follows: in section 2 we describe our methodology and present the results in section 3. Section 4 is destined to answer our Research Questions and we develop our conclusions in section 5.

## II. RELATED WORK

Concerns with Security Threats in Cloud Computing are quite recent, more precisely from 2008. In the last years the threats are receiving much attention by several researchers. In 2010, Farrell [12] alerts about problems of governance, risk and compliance of CC. In 2011, Hori et al [13] reports about security aspects for internal threats on CC. In the same year, Khorshed et al [14] propose two contributions: research in literature with focus on lacunae and challenges of threats, and defines an approach to prevention of attacks. In 2012, Ayala et al [15] identifies the threats and attacks and proposes solutions based in guides from NIST [16] and CSA [17]. In the same year, Yeluri et al [18] reports about experiences of Intel team with threats to security and resources control in CC. Also in 2012, Aqrabi et al [19] through a revision of literature and results obtained in simulations, proposes to identify the quests in adoption of security and compliance in CC. Nowadays, the theme of security threats in Cloud Computing is being well explored. We identified that 38% and 31% of the publications that we cataloged were made in 2011 and 2012, respectively, according to Figure 1. We were motivated to produce this work because, among the publications related to threats in current literature, we did not find one considering the type of solution proposed by the authors, that could identify which compliances were related in those publications.

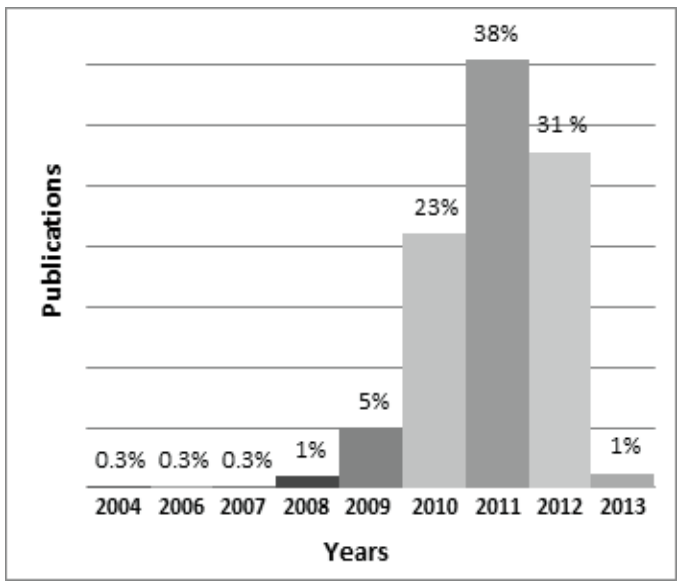

Figure 1. Percentage of publications spread by year.

## III. SYSTEMATIC MAPPING

### A. Research Questions

In this work we followed the orientations of Kitchenham [6] for elaboration of 4 Research Questions (RQ), with the goal to determine the content and conception of the systematic revision. Our work aims to answer the following RQ: (1) Which Security Threats to Models of Clouding Computing are being more addressed in literature? (2) Which Security Domains are being more explored by the Threats? (3) Which types of Solutions are being proposed in those approaches? (4) Among those approaches, which Compliances are involved?

### B. Definition of Research and Primary Source

We defined Elsevier Scopus as the primary source of our work. Besides having a considerable number of publications, we observed that in Scopus a large number of indexations of works from other sources are available. In addition, its search engine is able to be refined with several functionalities in its filters. Other sources chosen were: IEEExplore, ACM Digital Library, SpringerLink, Science Direct and Engineering Village.

As initial research, we searched in Scopus for works related to security on CCM, using the following filter rule: {[(Non-compliance with security) OR (key-words for security)] AND [cloud computing solutions]} in the title or abstract or key-words in the article. Resulting on the following Search String: ("flaw" OR "risk" OR "threat" OR "vulnerabilit*" OR "unsafe" OR "untrust") AND ("security" OR "safe" OR "trust") AND ("cloud" OR "multi-tenan*" OR "*aas" OR "* as a service" OR "* as-a-service"). Adding the results of all research sources, the total amounted to 1011 publications. Many of the occurrences were not in the research context and a manual refining based in the results or then triage had to be performed. We did not want to refine too much the Search String, because there was the risk of any relevant publication being excluded, we rather choose to leave the Search String wide open, leaving the refining in charge of a more detailed manual inspection.

### C. Inclusion Criteria

From that, we started our triage process considering the following inclusion criteria:

- Security in Cloud Computing as the main theme.
- The publication should have some relationship with one of the seven threats.
- The publication should have a proposed solution.

### D. Exclusion Criteria

- Duplication of publication.
- Journals not accessible online.
- Publications with related threat, but not active in cloud computing.
- Publications that only bring a revision or approach, without a proposal of solution.

*E. Relevance Criteria*

- Papers with well detailed solution proposal;
- Papers that have some kind of proposal validation, with statistical data, experiment, etc;
- Papers focused in fulfilling some compliance;

*F. Screening of Publications*

Each researcher applied the triage in a superficial way, based on the abstract of publications. When it was detected that at least a threat was applied, and some solution identified as a contribution, the publication was already considered. We found that for our research there were two cases where the superficial process was not enough, the first case when the abstract was too short, the other when it was not possible to extract from the abstract the proposal solution as a contribution. Those publications were allocated in separate for a more detailed future evaluation where they will be analyzed in introduction or in other chapters of the publication. This triage resulted in 661 publications according to Figure 2.

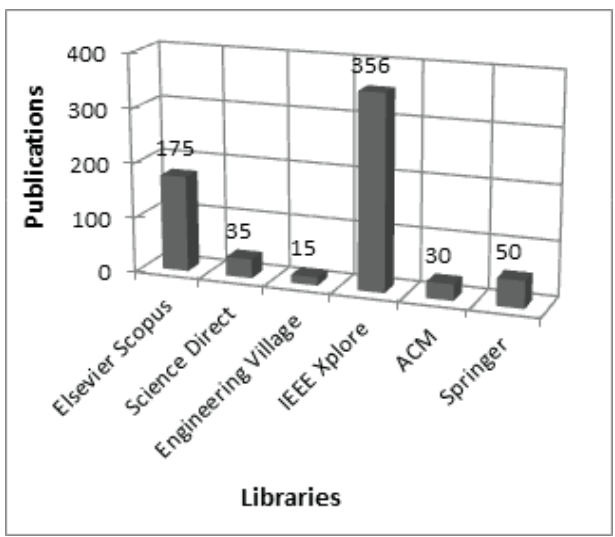

Figure 2. Publications spread by literary sources

*G. Results Classification*

In this phase, we made the analysis and classification of threats related to publications, the security domains involved, the type of solution proposed in each publication and if there is an approach aiming to fulfill some compliance standard. Posteriorly, the other authors interacted and analyzed the results related to the chosen classifications and reached the same conclusions.

## IV. ANALYSIS AND RESULTS DISCURSION

Here the questions of researches proposed in the protocol are answered.

*A. Result Obtained From RQ1:*

Despite each threat having a specific characteristic, nothing forbids it to act simultaneously with other threats in the same scene, resulting thus in several intersections between publications and threats. The seven threats are distributed in 661 publications according to Figure 3:

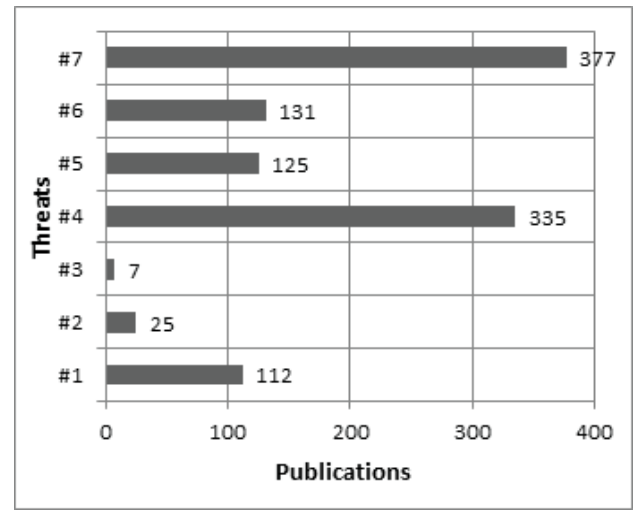

Figure 3. Distribution of publications by threats

*1) Threat #1: Abuse and Nefarious Use of Cloud Computing*

With 112 publications, it is an intermediary threat considering literary exploration. The consequence of this threat helps the growth of plagues like botnets, from which come problems like Distributed Denial of Service (DDoS), solves of Completely Automated Public Turing test to tell Computers and Humans Apart (CAPTCHA), storage of malicious files and botnet networks [3]. This threat evidences the fact that today it is very simple for any user to hire a cloud computing solution, it is even possible to get a free evaluation time, having only a valid credit card, which could come from a robbery or fraud. This ends up encouraging the action of malicious people to inject spam, malwares or even to practice illicit activities on the cloud [3]. There is only one proviso: until this moment, the version 2.0 of Top Threats Cloud Computing of CSA was not officially released, but a Survey was disclosed determining that instead of seven, there would be eight fails. This is because problems related to DDos are being so explored that it was dismembered and became a distinct threat in order to ease the understanding of strategies for its prevention [8].

*2) Threat #2: Insecure Interfaces and APIs*

A very relevant area, but so far not explored enough in literature. We cataloged only 25 papers. There are thousands of available APIs to be consumed, being also possible to build combinations of other APIs, known as Mashups. Those interfaces have serious standardization problems [4], this makes hard to apply a consistent security policy and the consequence is that many times access control, authentication, entry treatment, traffic of encrypted data, monitoring of activities, among other security aspects are neglected, offering a huge risk to cloud computing [3].

*3) Threat #3: Malicious Insiders*

This threat represents attacks of an active employee, ex-employee or business partner from the cloud provider that somehow has an authorized access and compromised the CIA of information stored in the cloud [10]. We consider this the hardest threat to be mitigated, having found only 7 publications related in literature. Despite being a very uncommon situation, its damage could be devastating [3], and it becomes even more critical when executed in environments without a straight access control of employees and without a well structured auditing that supports forensic analysis.

*4) Threat #4: Shared Technology Issues*

The second most explored threat in literature, with 335 publications found, focusing in IaaS models. Some components of this architecture were not projected for the scalability demanded from the model, making necessary to implement virtual machine monitoring to manage its resources [3]. Many times this layer does not have an adequate defense strategy and does not exert a good monitoring of network security. This is the scene where this kind of threat is more present.

*5) Threat #5: Data Loss or Leakage*

This threat happens when an exclusion, change or improper appropriation of some data in the cloud is made [3]. We considered this the most explored threat nowadays, because it represents a large number of the most recent publications. The cloud solutions for Storage and Bigdata are also having a strong growth. In consequence, the worry to provide CIA to data is also emerging; we found 125 publications in our research.

*6) Threat #6: Account or Service Hijacking*

Methods of phishing, fraud and vulnerability exploration, besides password credentials used in distributed ways, give amplitude to this problem [3]. The worry with kidnapping of accounts was the target of many cloud providers already consolidated in the market, such as Amazon [9]. There were found 131 publications in literature.

*7) Threat #7: Unknow Risk Profile*

It is the most explored threat in literature, with 377 selected publications. In cloud computing the abstraction regarding architecture details and maintenance responsibilities proportionate a greater security with obscurity by the cloud providers [3]. Details such as software version, failure fixes in order to avoid problems such as zero day, process that meet good security practices, among other aspects that are many times neglected by the cloud provider, falling in the large problem of the transparency of quality of service offered to its consumers.

*B. Result Obtained From RQ2:*

In this section we identified the Security Domains involved in each threat. We elaborated our classification based in the one proposed by [Mather et al, "Cloud Security and Privacy: A Enterprise Perspective on Risks and Compliance", O'Reilly Media; 1 edition, 2009], where eight different domains are enumerated. We classified them in a more granular way in order to get better visibility of results from our research, resulting in 15 Security Domains. The prevention measures of each threat may involve one or more domains, therefore also subject to intersections, according to the distribution displayed in Figure 4. Table 1 shows a brief description of each domain and the total amount of related publications.

TABLE I. FACET 1: SECURITY DOMAINS

| Domain | Description | Score |
|---|---|---|
| Access Control | Intervene in user access, from what the user access is granted or denied to a given datum or resource. Covers practices as for example Single-Sign-On (SSO) and Role-Based Access Control (RBAC). | 39 |
| Accountability | Ensures the quality of information with regard to possible and undesired behaviours of a system or infrastructure in the cloud. | 8 |
| Anonymity | Refers to traffic of public data, not allowing the same to be intercepted, warranting anonimousity in public or mixed clouds. | 3 |
| Applied Cryptography | Capacity of an emissor to make its data unreadable, with only the receptor being able to read the content. | 27 |
| Authentication | Verifies and validates a user identification. | 16 |
| Data or Database Protection | Technique for protection of informations stored either in bigdata or storage. | 130 |
| Digital Forensic | Technique of systematic inspections in computational resources in order to collect informations that may evidence a supposed crime committed. Presents itself as an excellent solution for problems related to inside threats in the cloud. | 5 |
| Identify Management | Is the management to establish and keep identity records applied to an access policy to each finality or service. | 20 |
| Integrity | Is the way to warrant that an information or behaviour cannot be changed by non-authorized people. | 21 |
| Intrusion Detection | Is the capacity to analyze a traffic or content that has the intention to compromise the integrity of a system or computational resource. | 22 |
| Formal Security Model | Overall, it is a scheme to specify and apply security policies. | 203 |
| Network Security | Guidelines to monitoring non-authorized or incorrect access in the network. | 96 |
| Privacy | It is the control of availability of a given information or resource in a public or shared environment. | 73 |
| Risk Analysis and Management | A set of policies to warrant that security processes happen in an efficient and continuous way over time. | 81 |
| Trust Model and Management | A set of policies that help to identify and estimate threats in a systemic way. | 50 |

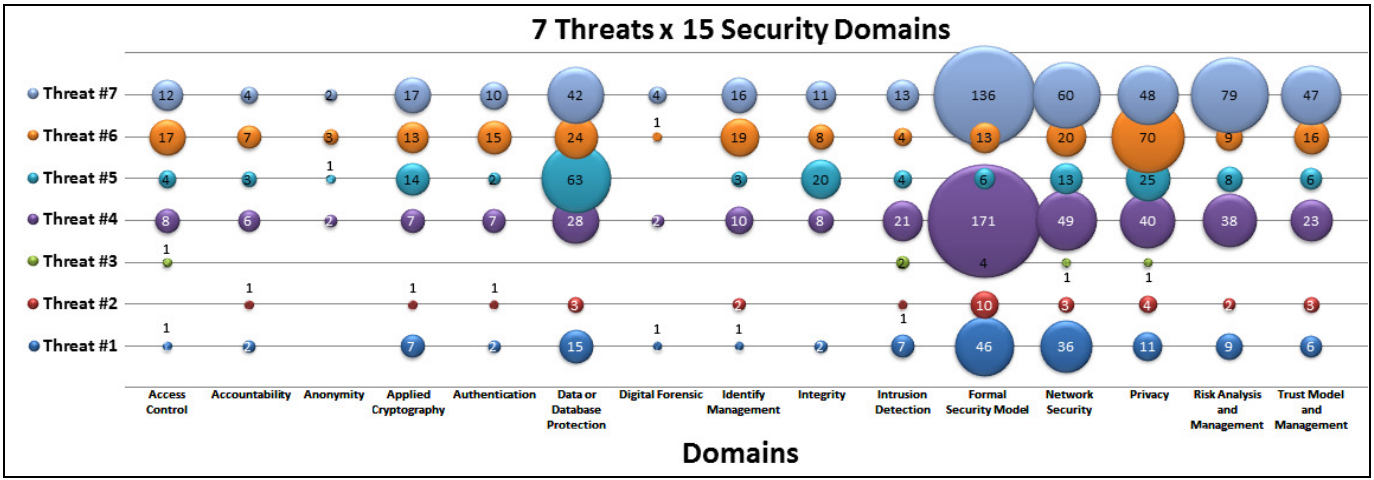

Figure 4. Distribution of publications by Threats and Domains

We detected the large number of combinations between "Access Control", "Data or Database Protection" and "Privacy" in more recent publications, and intrinsically linked to Threat #5. Many concerns involve techniques such as "Granular Access Control" and "Granular Audits" in the fields of Storage and Bigdata.

*C. Result Obtained From RQ3:*

In this stage we identified and classified the proposals of each work. Unlike the previous metrics, in this one there were no intersections. We assumed that each publication would have a single proposal, in cases when there was more than one, we considered the more elaborate by the authors. We measured eight types of proposals, according to Figure 5.

*D. Result Obtained From RQ4:*

In this stage we analyzed and identified, among the selected publications, those that were concerned with some compliance standard. Compliance is the condition of someone or of a group of people or processes to be according to what is desired or previously established, the desired in question are the specification standards. In this stage there were 9 intersections, as for example, we observed a publication that focus in compliances from NIST and FCAPS. As the answer for the RQ we identified a total of 18 compliances according to Figure 6.

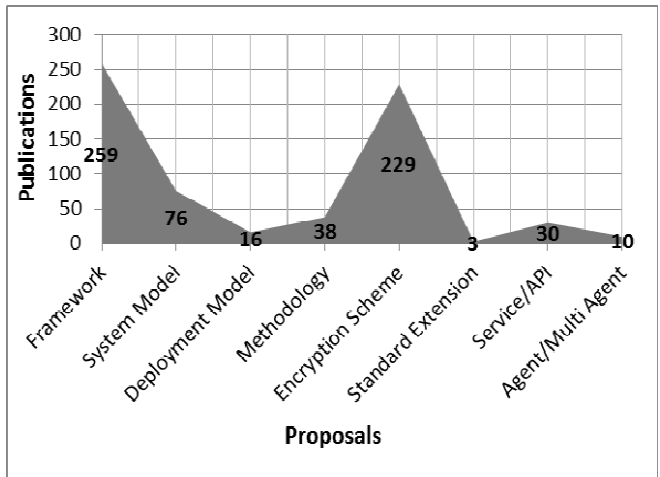

Figure 5. Distribution of publications by Proposals

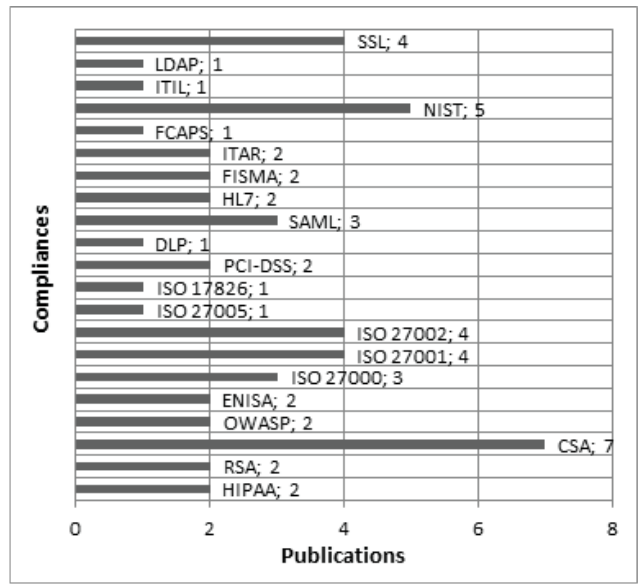

Figure 6. Distribution of publications by Compliances

*1) NIST*

Founded in 1901, is a non-governmental non-regulatory agency of USA trade. It has research in several areas, among them to promote standards to technological processes. After 15 revisions, it created the specification where it defines CC with 5 characteristics: self-service on demand, access to broadband network, resources pool, fast elasticity and measurable service. This definition was the milestone of standardization in the cloud. Other organizations such as CSA have this specification as the base for their researches [16].

*2) FCAPS*

It is an ISO standard, defining itself a model for network management composed by 5 levels: F: Fault, C: Configuration, A: Accounting, P: Performance and S: Security.

In level F are fixed the errors identified. It is also performed a management for prevention of future errors. In level C is performed a monitoring both in the network as well as in development changes. In this stage obsolete software or resources are removed from the network ecosystem and periodical updates of equipment and software are performed. Level A is dedicated to allocation and distribution of resources offered by the network, warranting that users receive resources according to the SLA. Level P is the management of performance where it is intended to identify problems and improvements. Level S is to ensure CID in all network resources [55].

*3) ITAR*

The International Traffic in Arms Regulation is a set of rules that control imports and exports of objects related to guns and ammunition [56].

*4) FISMA*

It is a federal law of the USA that recognizes the importance of information security in federal agencies data, demanding that each agency complies with security processes that control its assets. Compliance with FISMA was formalized by NIST in publication 800-53 [57].

*5) HL7*

Is a standard from the American National Standards Institute, ANSI, used to storage and handling of medical data. Any and all information related to patients, doctors and drugs is constructed from technical terms; this standard has the goal to universalize this communication [58].

*6) SAML*

It is a standard created by OASIS applied in the exchange of authentication and authorization of data between distinct security domains, based in protocols of Token exchange using XML, giving support to Web platforms and techniques such as SSO [59].

*7) DLP*

It is a technique to avoid, in time, incidents of violation or undue access to sensible data. The consequences may change, from access inhibition to the file or self-destruction of it [60].

*8) PCI-DSS*

It is a security pattern created by the Payment Card Industry Security Standards Council (PCI SSC), aimed to concerns with implementations in software that will do transactions with credit card. Its goal is to standardize the implementation and evaluate the providers of that software [61].

*9) ISO 17826*

It is a security standard of information published by the International Standards Organization (ISO) and by the International Electrotechnical Commission (IEC). Also called CDMI, it specifies the interface to storage and management of data in the cloud. This documentation is focused in developers or users of cloud storage [62].

*10) ISO 27005*

It is a standard for information security published by the ISO/IED. Its goal is to provide orientation for management of information about security risks [64].

*11) ISO 27002*

It is a standard for information security published by the ISO/IEC. It has the goal to establish directives and general principles to implement keep and improve the management of information in an organization [65].

*12) ISO 27001*

It is a standard for information security published by the ISO/IEC. The rule focuses on the concerns with implementation, monitoring, improvement and revision of a Management System of Information Security (MSIS) [11].

*13) ISO 27000*

Is a standard for information security published by ISO/IEC. The rule is a standard about good practices in Management of Information Security, which brings companies to the higher international level of excellence in Information Security. [27]

*14) ENISA*

The European Network and Information Security Agency is an agency from the European Union. The agency has the goal to contribute for the development of a culture of information and network security for the benefit of citizens, consumers, companies and public sector organizations of the European Union. In consequence, it will contribute for the good functioning of the internal marked of the European Union [20].

*15) OWASP*

The Open Web Application Security Project (OWASP) is an open source project for security of applications. The OWASP community has corporations, educational organizations and individuals from all over the world. This community works to create articles freely available, methodologies, documentation, tools and technologies that promote the good practices of security. The OWASP Foundation is a charity organization that supports and manages OWASP projects and its infrastructure. It is also a nonprofit registered trademark in Europe since June 2011 [51].

*16) CSA*

It is a nonprofit organization with the mission to promote the use of better practices, provide warranty of security in Cloud Computing, and provide education about cloud computing use to help to protect all kinds of computing. The Cloud Security Alliance is led by a wide coalition of industry professionals, companies, associations and other interested parties [3].

*17) RSA*

RSA is an algorithm for data cryptography, which owes its name to three teachers of the MIT (founders of the current company RSA Data Security, Inc.), Rivest, Shamir and Adleman. It is considered the most well succeeded implementation of asymmetric keys algorithms, and is based in classical theories of numbers. It was also the first algorithm to allow cryptography and digital signature, and one of the great inventions in public key cryptography [1].

*18) HIPAA*

It is the acronym for Insurance Portability and Accountability Act. It was approved by the American Congress in 1996, during the Bill Clinton government. It is a standard with the goal to protect data related to health, ensuring privacy and fraud prevention [63].

*E. Relevant Works*

The publications that we considered relevant are based in the criteria of relevance defined in the protocol. We selected three publications of threat #1, nine publications of threats #4 and #5, four publications of threat #6 and nine publications of threat #7, totaling 34 publications considered the most relevant in our research result. Curiously the result of our search reveals that none of the works related to threats #2 and #3 were concerned to fulfill any compliance.

*1) Threat #1*

TABLE II.  COMPLIANCES INTO THREATS #1

| Compliance | Proposal | Domain | Reference |
|---|---|---|---|
| ISO 27001 | Standard Extension | Formal Security Model | Ristov et al [38] |
| ITAR | Framework | Formal Security Model | Wang et al [53] |
| ITIL | Framework | Formal Security Model | Kamer & Vranken [25] |

*2) Threat #4*

TABLE III.  COMPLIANCES INTO THREATS #4

| Compliance | Proposal | Domain | Reference |
|---|---|---|---|
| ISO 27000, ISO 27001, ISO 27002. | Framework | Risk Analysis and Management | Zhao [36] |
| ISO 27001, ISO 27002. | Methodology | Authentication | Auty et al [39] |
| ISO 27001 | Framework | Formal Security Model | Julich & Hall [40] |
| ISO 27002 | Framework | Formal Security Model | Rebollo et al [41] |
| PCI-DSS | Framework | Trust Analysis and Management | Hizver & Chiueh [43] |
| PCI-DSS | System Model | Privacy | Kounelis et al [44] |
| HL7 | Deployment Model | Formal Security Model | Mouleeswaran et al [50] |
| ITAR | Framework | Formal Security Model | Poolsappasit et al [52] |
| NIST | System Model | Privacy | Kim et al [54] |

*3) Threat #5*

TABLE IV.  COMPLIANCES INTO THREATS #5

| Compliance | Proposal | Domain | Reference |
|---|---|---|---|
| HIPAA | Encryption Scheme | Access Control, Integrity Privacy, Applied Cryptography, Data or Database Protection | Li et al [28] |
| HIPAA | System Model | Access Control, Integrity Privacy, Data or Database Protection | Huemer et al [29] |
| RSA | Encryption Scheme | Formal Security Model, Applied Cryptography, Data or Database Protection | Saravanan et al [30] |
| RSA | Encryption Scheme | Formal Security Model, Applied Cryptography, Data or Database Protection | Lin et al [31] |
| ISO 17826 | Standard Extension | Formal Security Model | Teckelmann et al [42] |
| DLP | Encryption Scheme | Formal Security Model, Applied Cryptography, Data or Database Protection | Basak et al [45] |
| LDAP | Encryption Scheme | Formal Security Model, Applied Cryptography | Zissis & Lekkas [22] |
| SSL | Encryption Scheme | Applied Cryptography, Authentication, Data or Database Protection | Mansukhani & Zia [23] |
| SSL | Framework | Formal Security Model, Data or Database Protection | Ahmed et al [24] |

*4) Threat #6*

TABLE V. COMPLIANCES INTO THREATS #6

| Compliance | Proposal | Domain | Reference |
|---|---|---|---|
| SAML | Framework | Access Control, Authentication, Identify Management | Lonea et al [46] |
| SAML | Framework | Identify Management, Trust Model and Management | Cabarcos et al [47] |
| SAML | Encryption Scheme | Identify Management, Applied Cryptography | Guerrero et al [48] |
| HL7 | Encryption Scheme | Access Control, Risk Analysis and Management | Sharma et al [49] |

*5) Threat #7*

TABLE VI. COMPLIANCES INTO THREATS #7

| Compliance | Proposal | Domain | Reference |
|---|---|---|---|
| NIST, CSA | Methodology | Formal Security Model | Ayala et al [15] |
| NIST, FCAPS | Framework | Risk Analysis and Management | Sitaram & Manjunath [20] |
| NIST, FISMA | Framework | Risk Analysis and Management | Almorsy et al [21] |
| CSA, ENISA | Deployment Model | Risk Analysis and Management | Kao et al [26] |
| CSA | Framework | Identify Management, Risk Analysis and Management | Bhardwaj & Kumar [32] |
| CSA, OWASP | Framework | Risk Analysis and Management | Saripalli & Walters [33] |
| OWASP, ISO 27002 | Service/API | Risk Analysis and Management | Chou & Oetting [34] |
| ENISA | Framework | Risk Analysis and Management, Formal Security Model | Liu et al [35] |
| ISO 27000, ISO 27005 | Standard Extension | Risk Analysis and Management, Formal Security Model | Beckers et al [37] |

## V. CONCLUSION

Our work has the goal to catalog the state of the art of publications available in literature, that report approaches about security threats in CC. We hope to help researchers who want to engage in the field and want to propose some solution to those problems. With our protocol we identified 661 publications about the subject, where we can analyze the Security Domains involved. We also presented types of solutions proposed by the authors, and identified that some of those publication were concerned with the compliance of some standard. We presented those compliances and reference the respective publications to ease the work of the researcher that wants to explore a specific compliance. We identified that Threat #7 is the most explored in literature and, in consequence, the Domains of Risk Analysis and Management and Trust Model and Management have expressive results. We also identified many combinations of Domains related to Access Control, Applied Cryptography, Data or Database Protection and Privacy. This reflects in the recent growth of publications that report experiences in solutions for Storage and Bigdata in the cloud. In this same scenario, we identified that Framework and Encryption Scheme are the most used solutions. Regarding compliances, the most present in publications are those indicated by CSA, ISO 27002, ISO 27001 and NIST. However we also found some works where its authors propose the extension of an ISO standard to solve a given problem. For future works, we are planning to investigate in more detail the obstacles of a given compliance to be inserted in CC scene.

## VI. ACKNOWLEDGEMENTS

We would like to acknowledge all the anonymous referees for their very helpful comments and suggestions.